\documentclass[11pt,a4paper]{article}
\usepackage{jheppub_kim}

\usepackage{subfigure,amsmath,amssymb ,amsfonts, latexsym}

\usepackage[notcite,color,final
                               ]{showkeys}
\definecolor{refkey}{gray}{.25}
\definecolor{labelkey}{gray}{.25}

\newcommand{\Eqn}[1]{&\hspace{-0.2em}#1\hspace{-0.2em}&}

\newcommand{\be}{\begin{equation}}
\newcommand{\ee}{\end{equation}}
\newcommand{\beqa}{\begin{subequations}\begin{eqnarray}}
\newcommand{\eeqa}{\end{eqnarray}\end{subequations}}

\newcommand{\ud}{\,\mathrm{d}}

\title{Observational Constraints on Teleparallel Dark Energy}

\author[a,b,c]{Chao-Qiang Geng}
\author[a,c]{Chung-Chi Lee}
\author[d,b]{Emmanuel N. Saridakis}

\affiliation[a]{ Department of Physics, National Tsing Hua University,
Hsinchu, Taiwan 300}

\affiliation[b]{National Center for Theoretical Sciences, Hsinchu,
Taiwan 300}

\affiliation[c]{Kavli Institute for Theoretical Physics China,
Chinese Academy of Science, Beijing 100190}

\affiliation[d]{CASPER,
Physics Department,
Baylor University,
Waco, TX  76798-7310, USA}

\emailAdd{geng@phys.nthu.edu.tw}
\emailAdd{g9522545@oz.nthu.edu.tw}
\emailAdd{Emmanuel$_-$Saridakis@baylor.edu}


\abstract{
We use data from  Type Ia Supernovae (SNIa), Baryon
Acoustic Oscillations (BAO), and Cosmic Microwave Background
(CMB) observations  to constrain  the recently proposed teleparallel
dark energy scenario based on the teleparallel equivalent of
General Relativity, in which one adds a canonical scalar field, allowing
also for a nonminimal coupling with gravity. Using the power-law, the
exponential and the inverse hyperbolic cosine potential ansatzes, we show
that the scenario is compatible with observations. In particular, the data
favor a nonminimal coupling, and although the scalar field is canonical
the model can describe both the quintessence and phantom regimes.
}

\begin{document}

\maketitle

\section{Introduction.}

Recent cosmological observations
\cite{observc1,observc11,observc12,observc13,observc14}  indicate
that the observable universe experiences an accelerated expansion. Although
the simplest way is the consideration of a cosmological constant,
there are two alternative ways to explain this behavior. The first is to
modify the content of the universe by introducing the dark energy sector,
which can be based on a canonical scalar field (quintessence)
\cite{quintessence,quintessence1,quintessence2,quintessence3,quintessence4,
quintessence5,nonminimal,nonminimal1,nonminimal2,nonminimal3,nonminimal4,
gr-qc/0001066,nonminimal6,nonminimal7,astro-ph/0606287,nonminimal8,nonminimal9}, a phantom field
\cite{phant,phant1,phant2,phant3,phant4,phant5}, or on the
combination of both these fields in a unified scenario called quintom
\cite{quintom,quintom1,quintom2,quintom3,quintom4,quintom4p,quintom5} (see
\cite{Copeland:2006wr} for a review). The
other approach is to modify the gravitational sector itself (see
\cite{Nojiri:2006ri} for a review and references therein).

The {\em teleparallel} dark energy is a recently proposed scenario that
tries to incorporate the dark energy sector \cite{Geng:2011aj}. It is based
on the ``teleparallel'' equivalent of General Relativity (TEGR), that is on
its torsion instead of curvature formulation \cite{ein28,Hayashi79}, in
which one adds a canonical scalar field, and the dark energy sector is
attributed to this field. In the case where the field is minimally coupled
to gravity the scenario is completely equivalent to the standard quintessence
\cite{quintessence,quintessence1,quintessence2,quintessence3,quintessence4,
quintessence5}, both at the background and perturbation levels.
However, in the case where one switches on the nonminimal coupling between
the field and the torsion scalar, that is the only suitable gravitational
scalar in TEGR, the resulting scenario has a richer structure, exhibiting
quintessence-like or phantom-like behavior, or experiencing the
phantom-divide crossing \cite{Geng:2011aj}. The richer structure is
manifested in the absence of a conformal transformation to an equivalent
minimally-coupled model with transformed field and potential, which is
known to be able to describe only the quintessence regime.

Since the teleparallel dark energy exhibits interesting cosmological behavior,
in the present work we use observations in order to constrain the
parameters of the model. In particular, we use data from Type Ia
Supernovae (SNIa), Baryon Acoustic Oscillations (BAO), and Cosmic Microwave
Background (CMB) to plot likelihood-contours for the present
dark-energy equation-of-state, matter density
parameter and nonminimal coupling parameters, respectively.

The paper is organized as follows:
In Section \ref{TEGRquint} we
present the scenario of the teleparallel dark energy and  derive the relevant
equations for the cosmological evolution. In Section \ref{constraints} we
use observational data  to produce likelihood-contours of the
model parameters. Finally, Section \ref{Conclusions} is devoted to the
summary of our results.

\section{Teleparallel Dark Energy}
\label{TEGRquint}

Let us briefly review the teleparallel dark energy. As we stated in 
Introduction, it is based on the ``teleparallel'' equivalent of General
Relativity (TEGR)~\cite{ein28,Hayashi79}, in which instead of using
the torsionless Levi-Civita connection one uses the
curvatureless Weitzenb{\"o}ck one. The dynamical objects are the four
linearly independent vierbeins (these are parallel vector fields, referred to as
the appellations ``teleparallel'' or
``absolute parallelism'').
It is interesting to note that the torsion tensor is formed solely from products of the first
derivatives of the tetrad.

In particular, the vierbein field ${\mathbf{e}_A(x^\mu)}$ forms an
orthonormal basis for the tangent
space at each point $x^\mu$, that is $\mathbf{e}
_A\cdot%
\mathbf{e}_B=\eta_{AB}$, where $\eta_{AB}=diag (1,-1,-1,-1)$,
and furthermore the vector $\mathbf{e}_A$ can be analyzed with the use of
its components $e_A^\mu$ in a coordinate basis, that is
$\mathbf{e}_A=e^\mu_A\partial_\mu $
\footnote{We follow the notation of
\cite{Chen001,Dent001}, that is Greek indices $\mu, \nu,$... and capital
Latin
indices $A, B, $... run over all coordinate and tangent
space-time 0, 1, 2, 3, while lower case Latin indices (from the middle of
the alphabet) $i, j,...$ and lower case Latin indices (from the beginning
of the alphabet) $a,b, $... run over spatial and tangent
space coordinates 1, 2, 3, respectively. Finally,
we use the metric signature $(+,-,-,-)$.}.
In such a construction, the metric tensor is obtained from the
dual vierbein
as
\begin{equation}  \label{metrdef}
g_{\mu\nu}(x)=\eta_{AB}\, e^A_\mu (x)\, e^B_\nu (x).
\end{equation}
Consequently, the  torsion tensor of the Weitzenb%
\"{o}ck connection $\overset{\mathbf{w}}{\Gamma}^\lambda_{
\nu\mu}$ \cite{Weitzenb23} reads
\begin{equation}  \label{torsion2}
{T}^\lambda_{\:\mu\nu}=\overset{\mathbf{w}}{\Gamma}^\lambda_{
\nu\mu}-%
\overset{\mathbf{w}}{\Gamma}^\lambda_{\mu\nu}
=e^\lambda_A\:(\partial_\mu
e^A_\nu-\partial_\nu e^A_\mu),
\end{equation}
where $\overset{\mathbf{w}}{\Gamma}^\lambda_{\nu\mu}\equiv e^\lambda_A\: \partial_\mu
e^A_\nu$.

In the present formalism, all the information
concerning the
gravitational field is included in the torsion tensor
${T}^\lambda_{\:\mu\nu} $.
As described in \cite{Hayashi79}, the  ``teleparallel Lagrangian'' can
be constructed from this torsion tensor under the assumptions of invariance
under general coordinate transformations, global Lorentz transformations,
and the parity operation, along with requiring the Lagrangian density
to be the second order in the torsion tensor. In particular, it is the torsion
scalar  $T$, given by~\cite{ein28,Hayashi79,Maluf:1994ji,Arcos:2005ec}:
{\small{
\begin{equation}  \label{telelag}
\mathcal{L}=T=\frac{1}{4}T^{\rho
\mu \nu }T_{\rho \mu \nu }+\frac{1}{2}T^{\rho \mu \nu
}T_{\nu \mu \rho }-T_{\rho \mu }^{\ \ \rho }T_{\ \ \ \nu }^{\nu
\mu }.
\end{equation}}}
The simplest action in a universe governed by teleparallel
gravity is
\begin{eqnarray}  \label{action}
S =\int d^4x e
\left[\frac{T}{2\kappa^2}+\mathcal{L}_m\right],
\end{eqnarray}
where $e = \text{det}(e_{\mu}^A) = \sqrt{-g}$ (one could also include a
cosmological constant).
Variation with respect to the vierbein fields provides  equations of
motion, which are exactly the same as those of GR for every geometry
choice, and that is why the theory is called ``teleparallel equivalent to
General Relativity''.

In principle one has two ways of generalizing the action (\ref{action}),
inspired by the corresponding procedures of the standard General Relativity.
The first is to replace $T$ by an arbitrary function $f(T)$ 
\cite{BengocheafT,Linder:2010py,Myrzakulov:2010vz,Chen001,Wu001,
Bamba:2010iw,
Dent001,Zheng:2010am,Bamba:2010wb,Wang:2011xf,Yerzhanov:2010vu,Yang:2010ji,
Wu:2010mn,Bengochea001,Wu:2010xk,Zhang:2011qp,Ferraro001,Cai:2011tc,
Chattopadhyay001,Sharif001,Wei001,Ferraro002,Miao003,Boehmer004,Wei005,
Capozziello006,Wu007,Daouda001,Bamba:2011pz}, similar to $f(R)$
extensions of GR, and obtain new interesting terms in the field equations.
The other, on which we focus in the present work, is to add a canonical
scalar field in (\ref{action}), similar to the GR quintessence, allowing for
a nonminimal coupling between it and gravity.
This field will constitute the dark energy sector, and thus the
corresponding scenario is called ``teleparallel dark energy''
\cite{Geng:2011aj}.
In particular, the action will simply read:
\begin{equation}
S=\int\ud^{4}x e\Bigg[\frac{T}{2\kappa^{2}}
+ \frac{1}{2} \Big(\partial_{\mu}\phi\partial^{\mu}\phi+\xi
T\phi^{2}\Big) - V(\phi)+\mathcal{L}_m\Bigg]. \label{action2}
\end{equation}
Concerning the nonminimal coupling we emphasize that 
the
nonminimal coupling will be between the torsion and the scalar field (similar to
the standard nonminimal quintessence where the scalar field couples to the
Ricci scalar).

Variation of the action (\ref{action2}) with respect to the vierbein fields
yields the equations of motion
\begin{eqnarray}\label{eom2}
\left(\frac{2}{\kappa^2}+2 \xi
\phi^2 \right)\left[e^{-1}\partial_{\mu}(ee_A^{\rho}S_{\rho}{}^{\mu\nu} )
-e_{A}^{\lambda}T^{\rho}{}_{\mu\lambda}S_{\rho}{}^{\nu\mu}
-\frac{1}{4}e_{A}^{\nu
}T\right]\nonumber\\
-
e_{A}^{\nu}\left[\frac{1}{2}
\partial_\mu\phi\partial^\mu\phi-V(\phi)\right]+
  e_A^\mu \partial^\nu\phi\partial_\mu\phi\ \ \nonumber\\
+ 4\xi e_A^{\rho}S_{\rho}{}^{\mu\nu}\phi
\left(\partial_\mu\phi\right)
=e_{A}^{\rho}\overset {\mathbf{em}}T_{\rho}{}^{\nu},~~~~~~~~~~~~
\label{eom2}
\end{eqnarray}
where
$\overset{\mathbf{em}}{T%
}_{\rho}{}^{\nu}$ stands for the usual
energy-momentum tensor. 
Therefore, 
for a flat
Friedmann-Robertson-Walker (FRW) background metric
\begin{equation}
\label{FRWmetric}
ds^2= dt^2-a^2(t)\,\delta_{ij} dx^i dx^j
\end{equation}
and 
 a vierbein choice of the form
\begin{equation}  \label{FRWvierbeins}
e_{\mu}^A=\mathrm{diag}(1,a,a,a),
\end{equation}
where $t$ is the cosmic time, $x^i$ are the comoving spatial
coordinates and $a(t)$ is the scale factor,
we obtain the corresponding Friedmann equations:
\begin{eqnarray}
\label{FR1}
&&
H^{2}=\frac{\kappa^2}{3}\Big(\rho_{\phi}+\rho_{m}\Big),
\\
\label{FR2}
&&
\dot{H}=-\frac{\kappa^2}{2}\Big(\rho_{\phi}+p_{\phi}+\rho_{m}+p_{m}
\Big),~~~~
\end{eqnarray}
where $H=\dot{a}/a$ is the Hubble parameter and a dot denotes
differentiation with respect to $t$.
 In these expressions,
 $\rho_m$ and $p_m$ are the matter energy density and
 pressure, respectively, following the standard evolution equation
$\dot{\rho}_m+3H(1+w_m)\rho_m=0$, with $w_m= p_m/\rho_m$
the matter equation-of-state parameter. Additionally, we have introduced
the energy density and pressure of the scalar field
\begin{eqnarray}
\label{telerho}
 &&\rho_{\phi}=  \frac{1}{2}\dot{\phi}^{2} + V(\phi)
-  3\xi H^{2}\phi^{2},\\
 &&p_{\phi}=  \frac{1}{2}\dot{\phi}^{2} - V(\phi) +   4 \xi
H \phi\dot{\phi}
 + \xi\left(3H^2+2\dot{H}\right)\phi^2.\ \ \ \ \
 \label{telep}
\end{eqnarray}
Moreover, variation of the action with respect to the scalar field
provides its evolution equation, namely:
\begin{equation}
\ddot{\phi}+3H\dot{\phi}+6\xi H^2\phi+   V'(\phi)=0.
\label{fieldevol2}
\end{equation}
Note that in the above expressions we have used the useful relation
$T=-6H^2$, which straightforwardly arises from the calculation of
(\ref{telelag})
for the flat FRW geometry.

In this scenario, similar to the standard quintessence, dark
energy is attributed to the scalar field, and thus its equation-of-state
parameter ($w_{DE}$) reads:
\begin{equation}\label{EoS}
w_{DE}\equiv w_\phi=\frac{p_\phi}{\rho_\phi}.
\end{equation}
 As a result, one can see that the scalar field evolution (\ref{fieldevol2})
 leads to the standard relation
\begin{equation}\dot{\rho}_\phi+3H(1+w_\phi)\rho_\phi=0.
\end{equation}

The teleparallel dark energy proves to exhibit a very interesting cosmological
implication~ \cite{Geng:2011aj,Wei:2011}. In the minimally-coupled case the cosmological equations coincide
with those of the standard quintessence, both at the background and
perturbation levels. However, when the nonminimal
coupling is switched on, one can obtain a dark energy sector being
quintessence-like, phantom-like, or experiencing the phantom-divide
crossing during evolution, a behavior that is much richer comparing to
General Relativity (GR) with a scalar field \cite{Geng:2011aj}. Therefore,
it is
both interesting and necessary to use observations  to constrain
the parameters of the scenario. This is performed in the next section.

\section{Observational Constraints}
\label{constraints}

We use  Type Ia Supernovae (SNIa) from
the Supernova Cosmology Project (SCP) Union2 compilation
\cite{Amanullah:2010vv}, Baryon Acoustic Oscillations (BAO) data from the
Two-Degree Field Galaxy Redshift Survey (2dFGRS) and the Sloan Digital Sky
Survey Data Release 7 (SDSS DR7)~\cite{Percival:2009xn}, and the Cosmic
Microwave Background (CMB) radiation data from Seven-Year Wilkinson
Microwave Anisotropy Probe (WMAP) observations~\cite{Komatsu:2010fb}
 to examine the teleparallel dark energy scenario. Since the
model includes the scalar-field potential, we perform our analysis for
three different potential cases, given by:

\begin{itemize}
\item {\underline{Power-Law potential}}

This potential class is common in cosmology
\cite{powerlaw,powerlaw1,powerlaw2,powerlaw3,powerlaw4}. 
Although we
can
straightforwardly perform our analysis for an arbitrary exponent, for
simplicity we focus on the most interesting quartic case
\begin{equation}\label{powerlaw}
V(\phi)=V_0 \phi^4.
\end{equation}

\item {\underline{Exponential potential}}

Exponential potentials are also very common in the literature
\cite{expon,expon1,expon2,expon3},
which are necessary to be considered in 
 every observational constraining
analysis. In the following we use the ansatz
\begin{equation}
\label{exponential}
V(\phi)=V_0 e^{-\kappa\lambda\phi}.
\end{equation}

\item {\underline{Inverse  hyperbolic cosine  potential}}

The use of  hyperbolic cosine  potential, or power-law functions of it, has
also many cosmological implications \cite{hyperb,hyperb1,hyperb2}.
Although we
could perform our analysis for an arbitrary exponent, in the following we
focus on the inverse case, namely:
\begin{equation}
\label{inversecosh}
V(\phi)=\frac{V_0}{\cosh(\kappa\phi)}.
\end{equation}
\end{itemize}

We examine the constraints on the model parameters and the present values
of the density parameters, following the $\chi^{2}$-method for
the recent observational data. The detailed analysis method for
SNIa, BAO and CMB data is summarized in  Appendix.
In general, we are interested in producing the likelihood contours for
physically-interesting parameters, namely the
present dark-energy equation-of-state parameter $w_{DE_0}$, the present
matter density parameter $\Omega_{m0}$ and the nonminimal coupling
parameter $\xi$. We mention here that $\xi$ must always be bounded
according to a physical constraint, namely it must lead to positive
$\rho_{DE}$ and $H^2$ in relations (\ref{FR1}) and (\ref{telerho}),
respectively. In practice, $\xi$ is found to be mainly negative (in our
convention), and only a small window of positive values is theoretically
allowed. In our analysis, for each of the three potentials, we fit three
parameters, namely $w_{DE_0}$, the dimensionless Hubble parameter $h$ and
$\Omega_{m0}$($\xi$), and then we draw the likelihood-contours for 
 $1\sigma$ and $2\sigma$ confidence levels.

\begin{figure}[!]
\begin{center}
\includegraphics[width=7.2cm]{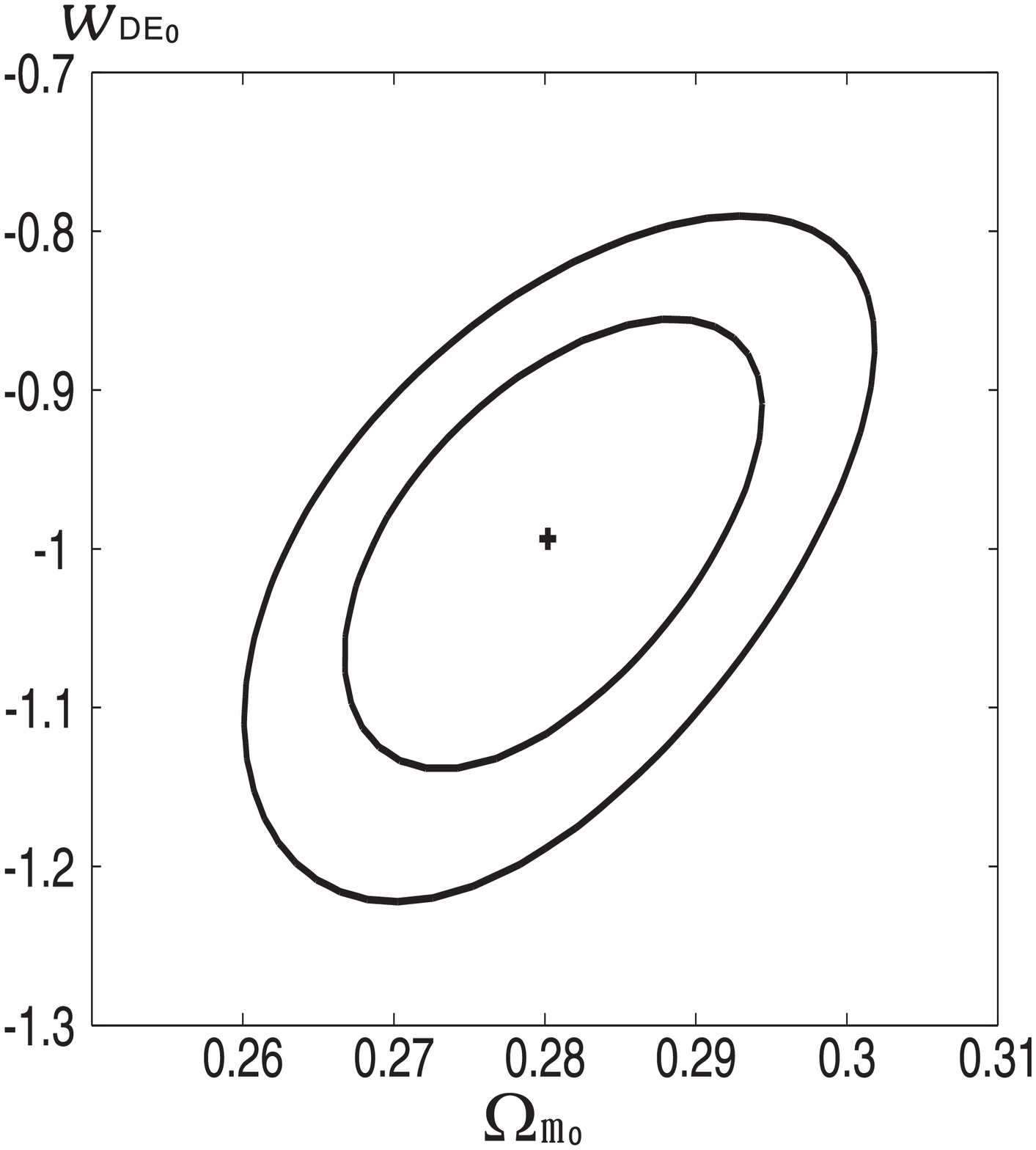}
\caption{{\it{ Contour plots of the present dark-energy
equation-of-state parameter $w_{DE_0}$ versus the present matter
density parameter $\Omega_{m0}$ under SNIa, BAO and CMB
observational data in the teleparallel dark energy scenario with the
quartic potential $V(\phi)=V_0 \phi^4$. The curves correspond 
to 1$\sigma$ and 2$\sigma$ confidence levels, respectively, and the cross
marks the best-fit point.}} }
\label{phi41}
\end{center}
\end{figure}

In Fig. \ref{phi41} we present the likelihood contours for $w_{DE_0}$  and
$\Omega_{m0}$ with the teleparallel dark energy scenario under the quartic
potential (\ref{powerlaw}).
As we observe, the scenario at hand is in agreement with observations, and
as expected, it can describe both the quintessence and phantom regimes.
Since the scalar field is canonical, it is a great advantage of
the present model.

\begin{figure}[!]
\begin{center}
\includegraphics[width=7.2cm]{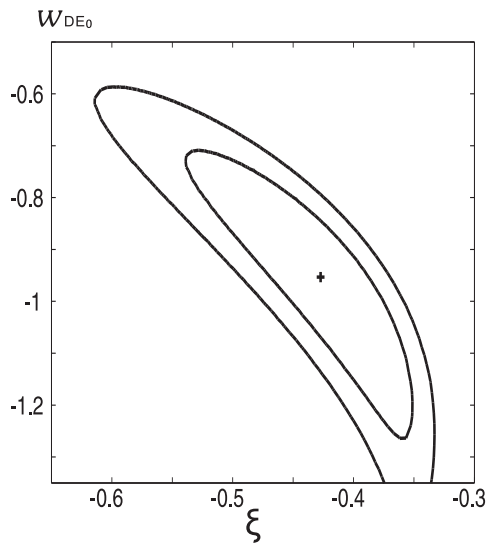}
\caption{{\it{ Contour plots of the present dark-energy
equation-of-state parameter $w_{DE_0}$ versus the nonminimal coupling
parameter $\xi$ under SNIa, BAO and CMB
observational data, in the teleparallel dark energy scenario with the
quartic potential $V(\phi)=V_0 \phi^4$. The curves correspond
to 1$\sigma$ and 2$\sigma$ confidence levels, respectively, and the cross
marks the best-fit point.}} }
\label{phi42}
\end{center}
\end{figure}

In Fig. \ref{phi42} we present the likelihood contours for $w_{DE_0}$  and
the nonminimal coupling parameter $\xi$, for the quartic potential
(\ref{powerlaw}).
Interestingly enough we observe that the nonminimal coupling is favored
by the data, and in particular a small $\xi$ is responsible for the
quintessence regime, while a larger one leads to the phantom regime. Note that
the best-fit values of $w_{DE_0}|_{b.f}\approx-0.98$ 
is very close to the
cosmological constant.

\begin{figure}[!]
\begin{center}
\includegraphics[width=7.2cm]{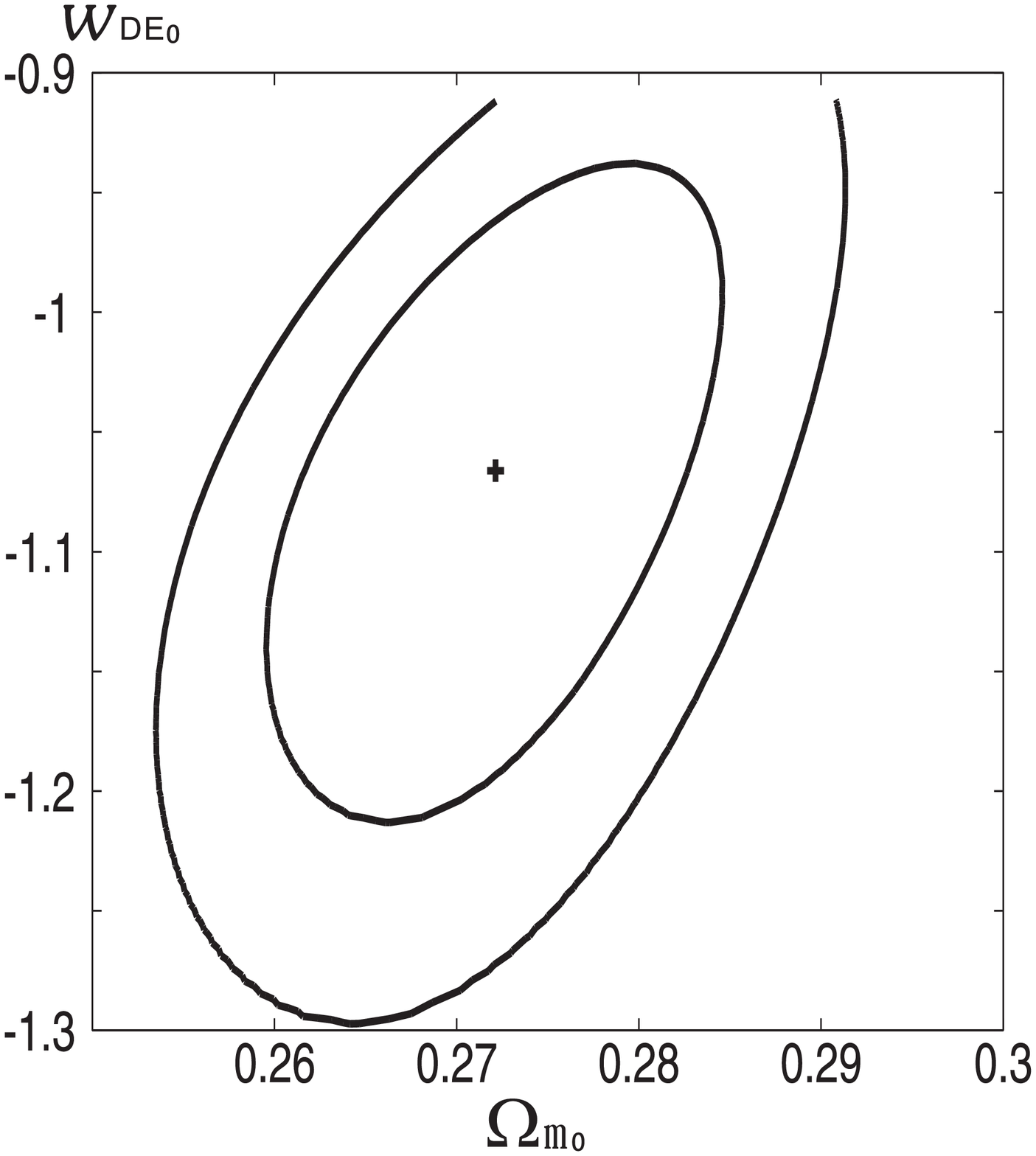}
\caption{{\it{Legend is the same as Fig.~\ref{phi41} but with
the exponential potential $V(\phi)=V_0 e^{-\kappa\lambda\phi}$. 
}} }
\label{exp1}
\end{center}
\end{figure}

In Fig. \ref{exp1} we present the likelihood contours for $w_{DE_0}$  and
$\Omega_{m0}$, for the teleparallel dark energy scenario under the
exponential potential (\ref{exponential}).
\begin{figure}[!]
\begin{center}
\includegraphics[width=7.2cm]{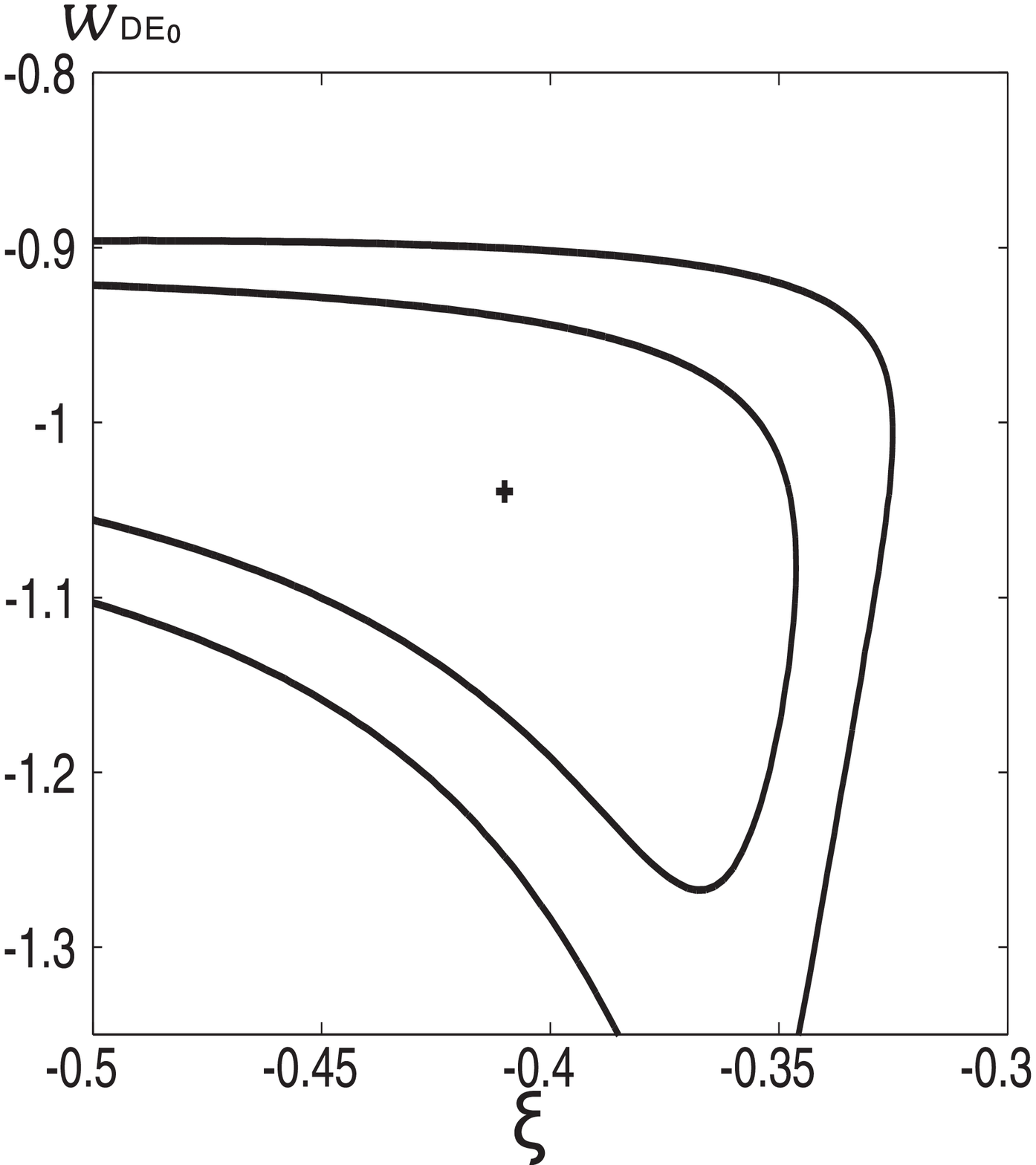}
\caption{{\it{ 
Legend is the same as Fig.~\ref{phi42} but with
the exponential potential $V(\phi)=V_0 e^{-\kappa\lambda\phi}$.
}} }
\label{exp2}
\end{center}
\end{figure}
As we observe, this scenario  is consistent with observations, and it
can describe both the quintessence and phantom regimes, with the phantom
regime favored by the data. Furthermore, in Fig. \ref{exp2} we present the
likelihood contours for $w_{DE_0}$  and
 $\xi$, for the exponential potential (\ref{exponential}).
From this graph we deduce that a non-minimal coupling is favored by the
data, and we observe that $w_{DE_0}$-values close to the cosmological
constant bound, either above or below it, can be induced by a relative
large $\xi$-interval, which is an advantage of this scenario. It is interesting to
mention that the exponential potential was used as an explicit example in
\cite{Geng:2011aj}, and our current analysis verifies its results.

\begin{figure}[ht]
\begin{center}
\includegraphics[width=7.2cm]{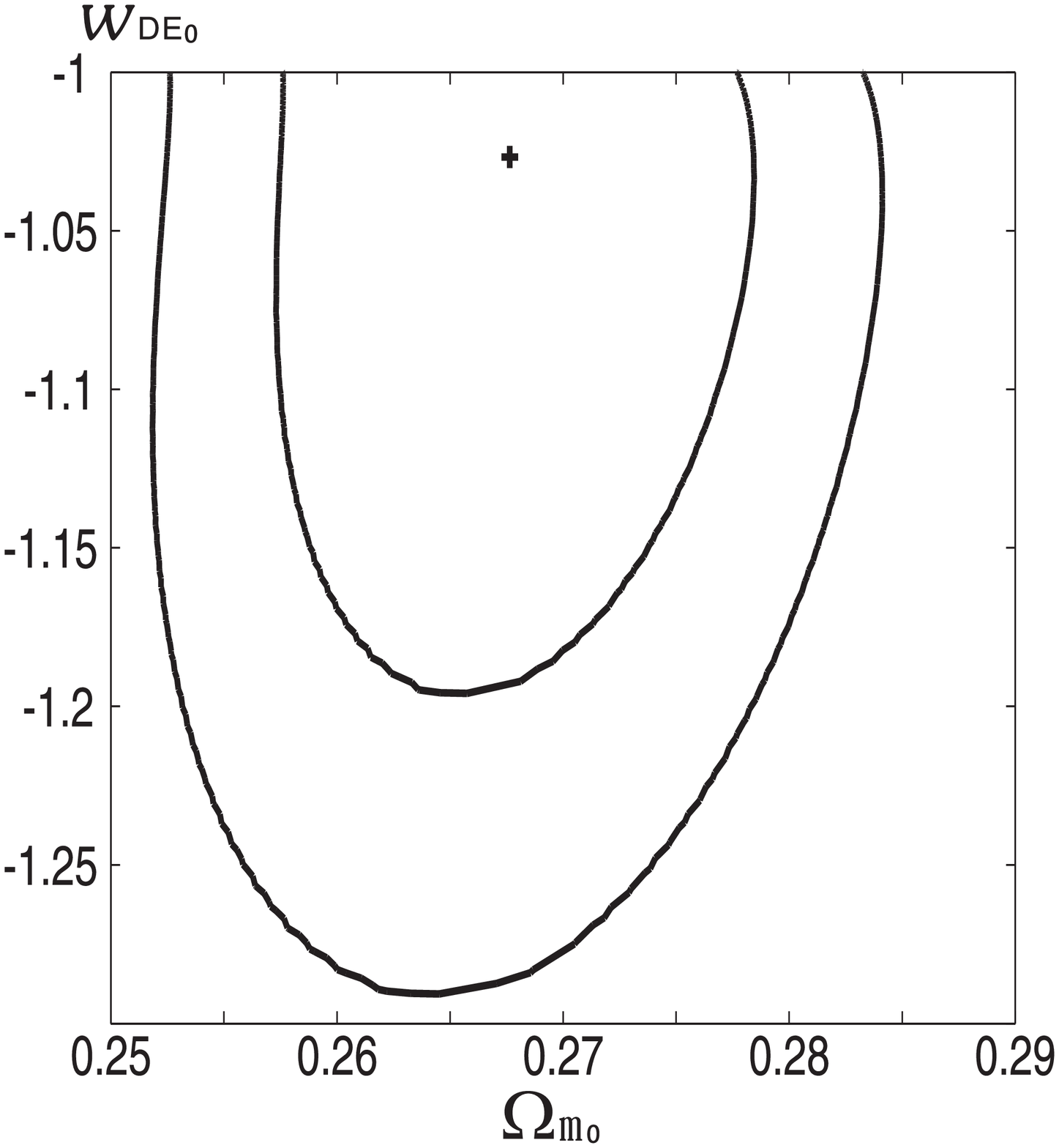}
\caption{{\it{
Legend is the same as Fig.~\ref{phi41} but with
%
the inverse  hyperbolic cosine  potential $V(\phi)=V_0 /
cosh(\kappa \phi)$. 
}} }
\label{cosh1}
\end{center}
\end{figure}

In Fig. \ref{cosh1} we depict the likelihood contours for $w_{DE_0}$  and
$\Omega_{m0}$, under the inverse  hyperbolic cosine 
potential (\ref{inversecosh}).
\begin{figure}[ht]
\begin{center}
\includegraphics[width=7.2cm]{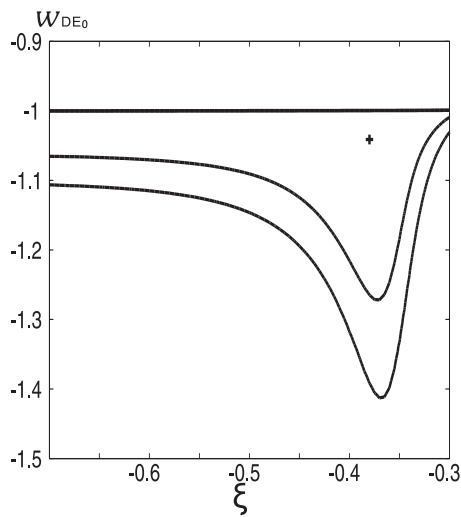}
\caption{{\it{ Legend is the same as Fig.~\ref{phi42} but with
%
the inverse  hyperbolic cosine  potential  $V(\phi)=V_0 / cosh(\kappa \phi)$. 
}} }
\label{cosh2}
\end{center}
\end{figure}
As we can see, this scenario is consistent with observations.
However, if we desire to avoid divergences in the $w_{DE}$ evolving
history, we are restricted in the phantom regime. In addition, the
best-fit value of  $w_{DE_0}|_{b.f}\approx-1.02$  is very close to the cosmological constant.
Furthermore, in Fig. \ref{cosh2} we present the
corresponding 
likelihood contours for $w_{DE_0}$ and $\xi$.
Similarly to the previous cases we can see that the nonminimal
coupling is favored by the data. It is interesting to note that $w_{DE_0}$-values close to
the cosmological constant bound can be induced by a relative
large $\xi$-interval.

We close this section with a comment on the positive values of the
nonminimal coupling $\xi$. As we mentioned above, the positivity
requirement for $\rho_{DE}$ and $H^2$ leads $\xi$ to be
negative, with only a small window of positive values theoretically
allowed. Now, in practice, if we perform our fitting procedure in the
positive $\xi$ region for the inverse  hyperbolic cosine  potential, we
find that positive $\xi$ is excluded. However, for the quartic and
exponential potentials we find the interesting result that for the
theoretically allowed positive $\xi$ ($0\leq\xi\lesssim 0.2$) 
$w_{DE}$
is always very close to a constant $w_{DE_0}$ with
$|w_{DE_0}-w_{DE}|\lesssim 10^{-3}$. The reason
 is that the scenario of the teleparallel dark energy for positive
$\xi$ (sufficiently small in order for the positivity of $\rho_{DE}$ and
$H^2$ not to be spoiled)  always results in the stabilization of $w_{DE_0}$ close
to the cosmological constant value, as can be proven by a detailed
phase-space analysis \cite{Inprep}. Such a behavior is an advantage
from both observational and theoretical point of view.
We 
would like to note for
comparison that, in the case of quintessence in the Einstein gravity and
in the presence of a dust-like component (CDM and baryons), dark energy
with an exactly constant $w_{DE}>-1$ is possible for the inverse hyperbolic
sine potential in some power as was first independently shown in Refs.~\cite{astro-ph/9904398}
and~\cite{astro-ph/0003364}.

\section{Conclusions}
\label{Conclusions}

In the present work we have used observational data  to impose
constraints on the parameters of the teleparallel dark energy scenario 
\cite{Geng:2011aj}, which is based on the teleparallel equivalent of
General Relativity (TEGR), that is on its torsion instead of curvature
formulation \cite{ein28,Hayashi79}. In this model one adds a canonical
scalar field, in which the dark energy sector is attributed, allowing also
for a nonminimal coupling between the field and the torsion scalar. Thus,
although the minimal case is completely equivalent with the standard
quintessence, the nonminimal scenario has a richer structure,
exhibiting the quintessence-like or phantom-like behavior, or experiencing the
phantom-divide crossing \cite{Geng:2011aj}.

In particular, we have fitted data from Type Ia Supernovae (SNIa), Baryon Acoustic
Oscillations (BAO), and Cosmic Microwave Background (CMB) observations
 to constrain the present dark-energy equation-of-state parameter
$w_{DE_0}$, the present matter density parameter $\Omega_{m0}$ and the
nonminimal coupling parameter $\xi$. Furthermore, in order to be general,
for the scalar-field potential we have taken three ansatzes, namely the
power-law, the exponential and the inverse hyperbolic cosine ones.

For the power-law (quartic) potential we have seen that teleparallel dark energy
is compatible with observations and, as expected, it can describe both the
quintessence and phantom regimes. Additionally, we have shown that the negative
nonminimal coupling is favored by the data, and in particular a small $\xi$
is responsible for the quintessence regime, while a larger one leads to the
phantom regime. For the exponential potential we have demonstrated  that both the
quintessence and phantom regimes can be described, with the phantom
regime favored by the data. Moreover, we have found that a negative non-minimal
coupling is favored and we have observed that $w_{DE_0}$-values close to the
cosmological constant bound, either above or below it, can be induced by a
relative large $\xi$-interval, which is an advantage of this scenario.
For the inverse hyperbolic cosine potential we have shown that $w_{DE_0}$ is
restricted in the phantom regime, while $\xi$ is restricted to negative
values, with a relatively large $\xi$-interval being able to lead to
$w_{DE_0}$-values close to $-1$. 
We remark that
positive values of $\xi$ are excluded for the inverse hyperbolic cosine 
potential, while for the power-law and exponential ones
$w_{DE_0}$ is very close to $-1$.

In summary, the scenario of the teleparallel dark energy is compatible with
observations, for all the examined scalar-field potentials. Furthermore, 
although the scalar field is canonical, the model can describe both the
quintessence and phantom regimes. These features are an advantage
from both observational and theoretical point of view, and they make
the scenario at hand a good candidate for the description of nature.
Finally, the data favor a nonminimal coupling, and thus the model is
distinguishable from the standard quintessence, since the two scenarios are
equivalent only for the minimal coupling.

An interesting and necessary investigation would be to go beyond the
background analysis, and examine observables that arise at the perturbation
level, such are the growth of matter overdensities and the
gravitational-wave spectrum, since these could also clarify possible
Lorentz-violation problems that are not seen at the background level
\cite{fTLorinv0,fTLorinv1,fTLorinv2}. Since this study lies beyond the
scope of the present
manuscript, it is left for future investigation.

\begin{acknowledgments}
 This work was partially supported by National Center of Theoretical
Science and  National Science Council (NSC-98-2112-M-007-008-MY3) of R.O.C.
Two of us (CQG and CCL) would like to thank KITPC for the wonderful program
of ``Dark Matter and New Physics''.
\end{acknowledgments}

\appendix

\section{Analysis method for the observational data}

In this appendix, we explain the methods for the elaboration of
observational data from Type Ia Supernovae (SNIa),
Baryon Acoustic Oscillations (BAO) and Cosmic Microwave Background (CMB)
radiation. The $\chi^{2}$ of the combined observational data is given by
\begin{equation}
\chi^{2}=\tilde{\chi}_{\mathrm{SN}}^{2}+\chi_{\mathrm{BAO}}^{2}+\chi_{
\mathrm{CMB}}^{2}.
\label{eq:A.21}
\end{equation}
In our fitting procedure we use the simple $\chi^{2}$
method, rather than the Markov-chain Monte Carlo (MCMC) approach such as
CosmoMC~\cite{Lewis:2002ah}. In the following we describe the calculation
for the various   $\chi^{2}_i$ of each observational dataset
(for detailed explanations on the data analysis see e.g.
\cite{Li:2009jx,Yang:2010xq}).
\\

{\it{a. Type Ia Supernovae (SNe Ia)}}\\

 SNe Ia observations provide the information on the luminosity distance
$D_{L}$ as a function of the redshift $z$.
The theoretical distance modulus $\mu_{\mathrm{th}}$ is defined by
\begin{equation}
\mu_{\mathrm{th}}(z_{i})\equiv5\log_{10}D_{L}(z_{i})+\mu_{0}\,,
\label{eq:A.1}\nonumber
\end{equation}
where $\mu_{0}\equiv42.38-5\log_{10}h$, with
$h \equiv H_{0}/100/[\mathrm{km} \, \mathrm{sec}^{-1} \,
\mathrm{Mpc}^{-1}]$ \cite{Komatsu:2010fb}.
The Hubble-free luminosity
distance for the flat universe is described as
\begin{equation}
D_{L}(z)=\left(1+z\right)\int_{0}^{z}\frac{dz'}{E(z')}\,,
\label{eq:A.2}\nonumber
\end{equation}
where $E(z) \equiv H(z)/H_{0}$,
with
{\small{
\begin{equation}
\frac{H(z)}{H_{0}}=
\sqrt{\Omega_{\mathrm{m}}^{(0)}\left(1+z\right)^{3}
+\Omega_{\mathrm{r}}^{(0)}\left(1+z\right)^{4}
+\Omega_{\mathrm{DE}}^{(0)}\left(1+z\right)^{3\left(1+w_{
\mathrm { DE } }
\right)}
}\,.
\label{eq:A.3}\nonumber
\end{equation}}}
Here,
$\Omega_{\mathrm{r}}^{(0)}=\Omega_{\gamma}^{(0)}
\left(1+0.2271N_{\mathrm{eff}}\right)$,
where $\Omega_{\gamma}^{(0)}$ is the present fractional photon energy
density
and $N_{\mathrm{eff}}=3.04$ is the effective number of neutrino
species~\cite{Komatsu:2010fb}.
We mention that $H(z)$ is evaluated by using numerical solutions of
the Friedmann equation.

The $\chi^{2}$ of the SNe Ia data is given by
\begin{equation}
\chi_{\mathrm{SN}}^{2}=\sum_{i}\frac{\left[\mu_{\mathrm{obs}}(z_{i})-
\mu_{\mathrm{th}}(z_{i})\right]^{2}}{\sigma_{i}^{2}}\,,
\label{eq:A.4}
\end{equation}
where $\mu_{\mathrm{obs}}$ is the observed value of the distance modulus.
In the following, subscriptions ``th" and ``obs" denote
the theoretically predicted and observed values, respectively.
$\chi_{\mathrm{SN}}^{2}$ is minimized
with respect to $\mu_{0}$,
which relates to the absolute magnitude,
since the absolute magnitude of SNe Ia is not known.
$\chi_{\mathrm{SN}}^{2}$ in (\ref{eq:A.4}) is expanded as \cite{CS-SN}
\begin{equation}
\chi_{SN}^{2}=A-2\mu_{0}B+\mu_{0}^{2}C\,,
\label{eq:A.5}\nonumber
\end{equation}
with
\begin{eqnarray}
&&A
=
\sum_{i}\frac{\left[\mu_{\mathrm{obs}}(z_{i})-\mu_{\mathrm{th}}(z_{i};\mu_{
0}=0)\right]^{2}}{\sigma_{i}^{2}}\,,
\quad\nonumber\\
&&B
=
\sum_{i}\frac{\mu_{\mathrm{obs}}(z_{i})-\mu_{\mathrm{th}}(z_{i};\mu_{0}=0)}
{\sigma_{i}^{2}}\,,
\quad\nonumber\\
&&C=\sum_{i}\frac{1}{\sigma_{i}^{2}}\,.
\label{eq:A.6}\nonumber
\end{eqnarray}
Thus, the minimum of $\chi_{\mathrm{SN}}^{2}$ with respect to $\mu_{0}$ is
expressed
as
\begin{equation}
\tilde{\chi}_{\mathrm{SN}}^{2}=A-\frac{B^{2}}{C}\,.
\label{eq:A.7}
\end{equation}
In our analysis we apply expression (\ref{eq:A.7}) for
the $\chi^{2}$ minimization and we use
the Supernova Cosmology Project (SCP) Union2 compilation, which contains
557 supernovae~\cite{Amanullah:2010vv}, ranging from $z=0.015$ to $z=1.4$.
\\

{\it{b. Baryon Acoustic Oscillations (BAO)}}\vspace{0.5cm}

The distance ratio of $d_{z}\equiv r_{s}(z_{\mathrm{d}})/D_{V}(z)$
is measured by BAO observations, where $D_{V}$ is the volume-averaged
distance, $r_{s}$ is the comoving
sound horizon and $z_{\mathrm{d}}$ is the redshift
at the drag epoch~\cite{Percival:2009xn}.
The volume-averaged distance $D_{V}(z)$ is defined as~\cite{BAO}
\begin{equation}
D_{V}(z)\equiv\left[\left(1+z\right)^{2}
D_{A}^{2}(z)\frac{z}{H(z)}\right]^{1/3}\,,
\label{eq:A.8}\nonumber
\end{equation}
where $D_{A}(z)$ is the proper angular diameter distance for the flat
universe, defined by
\begin{equation}
D_{A}(z) \equiv
\frac{1}{1+z}\int_{0}^{z}\frac{dz'}{H(z')}\,.
\label{eq:A.9}\nonumber
\end{equation}
The comoving sound horizon $r_{s}(z)$ is given by
{\small{
\begin{equation}
r_{s}(z)=\frac{1}{\sqrt{3}}\int_{0}^{1/(1+z)}\frac{da}{a^{2}
H(z^{\prime}=1/a-1) \sqrt{1+\left(3\Omega_{b}^{(0)}/4\Omega_{\gamma}^{(0)}
\right)a}},
\label{eq:A.10}\nonumber
\end{equation}}}
where $\Omega_{b}^{(0)} = 2.2765 \times10^{-2} h^{-2}$ and
$\Omega_{\gamma}^{(0)} = 2.469\times10^{-5}h^{-2}$ are the current
values of baryon and photon density parameters,
respectively~\cite{Komatsu:2010fb}.
The fitting formula for $z_{\mathrm{d}}$ is given
by~\cite{Eisenstein:1997ik}
\begin{equation}
z_{\mathrm{d}}=\frac{1291(\Omega_{\mathrm{m}}^{(0)}h^{2})^{0.251}}
{1+0.659(\Omega_{\mathrm{m}}^{(0)}h^{2})^{0.828}}\left[1+b_{1}
\left(\Omega_{b}^{(0)}h^{2}\right)^{b2}\right]\,,
\label{eq:A.11}\nonumber
\end{equation}
with
\begin{eqnarray}
&&b_{1}=0.313(\Omega_{\mathrm{m}}^{0}h^{2})^{-0.419}\left[
1+0.607\left(\Omega_{\mathrm{m}}^{0}h^{2}\right)^{0.674}\right]\,, \quad
\nonumber\\
&&b_{2}=0.238\left(\Omega_{\mathrm{m}}^{0}h^{2}\right)^{0.223}\,.\nonumber
\label{eq:A.12}
\end{eqnarray}
The typical value of $z_{\mathrm{d}}$ is about $1021$
for $\Omega_{\mathrm{m}}^{(0)}=0.276$ and $h=0.705$.

According to the BAO data from
the Two-Degree Field Galaxy Redshift Survey (2dFGRS)
and the Sloan Digital Sky Survey Data Release 7
(SDSS DR7)~\cite{Percival:2009xn},
the distance ratio $d_{z}$ at two redshifts $z=0.2$ and
$z=0.35$
is measured to be $d_{z=0.2}^{\mathrm{obs}}=0.1905\pm0.0061$ and
$d_{z=0.35}^{\mathrm{obs}}=0.1097\pm0.0036$,
with the inverse covariance matrix:
%
\begin{equation}
C_{\mathrm{BAO}}^{-1}=\left(
\begin{array}{cc}
30124 & -17227\\
-17227 & 86977
\end{array}\right)\,.
\label{eq:A.13}\nonumber
\end{equation}
%
Finally, the $\chi^{2}$ for the BAO data is calculated as
%
\begin{equation}
\chi_{\mathrm{BAO}}^{2}=
\left(x_{i,\mathrm{BAO}}^{\mathrm{th}}-x_{i,\mathrm{BAO}}^{\mathrm{obs}}
\right)
\left(C_{\mathrm{BAO}}^{-1}\right)_{ij}
\left(x_{j,\mathrm{BAO}}^{\mathrm{th}}-x_{j,\mathrm{BAO}}^{\mathrm{obs}}
\right),
\label{eq:A.14}\nonumber
\end{equation}
%
where $x_{i,\mathrm{BAO}} \equiv \left(d_{0.2},d_{0.35}\right)$.
\\

{\it{c. Cosmic Microwave Background (CMB) radiation}}\\

The CMB observational data are sensitive to the distance
to the decoupling epoch $z_{*}$~\cite{Komatsu:2010fb}.
Hence, by using these data we obtain
constraints on the model in the high redshift regime ($z\sim1000$).

The acoustic scale $l_{A}$ and
the shift parameter
$\mathcal{R}$~\cite{Bond:1997wr}
are defined by
\begin{eqnarray}
l_{A}(z_{*})
\Eqn{\equiv}
\left(1+z_{*}\right)\frac{\pi D_{A}(z_{*})}{r_{s}(z_{*})}\,,
\label{eq:A.15}\nonumber \\
\mathcal{R}(z_{*})
\Eqn{\equiv}
\sqrt{\Omega_{\mathrm{m}}^{(0)}}H_{0}
\left(1+z_{*}\right)D_{A}(z_{*}),
\label{eq:A.16} \nonumber
\end{eqnarray}
where $z_{*}$ is the redshift of the decoupling epoch,
given by~\cite{Hu:1995en}
{\small{
\begin{equation}
z_{*}=1048\left[1+0.00124\left(\Omega_{b}^{(0)}h^{2}\right)^{-0.738}\right]
\left[1+g_{1}\left(\Omega_{\mathrm{m}}^{(0)}h^{2}\right)^{g2}\right],
\label{eq:A.17}\nonumber
\end{equation}}}
with
{\small{
\begin{equation}
g_{1}=\frac{0.0783\left(\Omega_{b}^{(0)}h^{2}\right)^{-0.238}}{
1+39.5\left(\Omega_{b}^{(0)}h^{2}\right)^{0.763}},\quad
g_{2}=\frac{0.560}{1+21.1\left(\Omega_{b}^{(0)}h^{2}\right)^{1.81}}.
\label{eq:A.18}\nonumber
\end{equation}}}
We use the
data from Seven-Year Wilkinson Microwave Anisotropy Probe (WMAP)
observations~\cite{Komatsu:2010fb} on CMB.

The $\chi^{2}$ of the CMB data is
\begin{equation}
\chi_{\mathrm{CMB}}^{2}=\left(x_{i,\mathrm{CMB}}^{\mathrm{th}}-x_{i,\mathrm
{CMB}}^{\mathrm{obs}}\right)
\left(C_{\mathrm{CMB}}^{-1}\right)_{ij}
\left(x_{j,\mathrm{CMB}}^{\mathrm{th}}-x_{j,\mathrm{CMB}}^{\mathrm{obs}}
\right),
\label{eq:A.19}\nonumber
\end{equation}
where $x_{i,\mathrm{CMB}}\equiv\left(l_{A}(z_{*}),
\mathcal{R}(z_{*}), z_{*}\right)$
and $C_{\mathrm{CMB}}^{-1}$ is the inverse covariance matrix.
The data from WMAP7 observations~\cite{Komatsu:2010fb}
lead to $l_{A}(z_{*})=302.09$,
$\mathcal{R}(z_{*})=1.725$ and $z_{*}=1091.3$ with the inverse covariance
matrix:
\begin{equation}
C_{\mathrm{CMB}}^{-1}=\left(\begin{array}{ccc}
2.305 & 29.698 & -1.333\\
29.698 & 6825.27 & -113.180\\
-1.333 & -113.180 & 3.414\end{array}\right)\,.
\label{eq:A.20}\nonumber
\end{equation}

\providecommand{\href}[2]{#2}

\begingroup

\raggedright

\endgroup

\end{document}